**Data-driven Tracking of the Bounce-back Path after Disasters: Critical Milestones of Population Activity Recovery and Their Spatial Inequality**

Yuqin Jiang[1,*], Faxi Yuan[1], Hamed Farahmand[1], Kushal Acharya[1], Jingdi Zhang[2], Ali Mostafavi[1]

[1] UrbanResilience.AI Lab, Zachry Department of Civil and Environmental Engineering, Texas A&M University, DLEB 802B, College Station, TX, 77843

[2] Department of Computer Science and Engineering, Texas A&M University, College Station, TX, 77843

*yuqinjiang@tamu.edu

**Abstract**
The ability to measure and track the speed and trajectory of a community's post-disaster recovery is essential to inform resource allocation and prioritization. The current survey-based approaches to examining community recovery, however, have significant lags and put the burden of data collection on affected people. Also, the existing literature lacks quantitative measures for important milestones to inform the assessment of recovery trajectory. Recognizing these gaps, this study uses location-based data related to visitation patterns and credit card transactions to specify critical recovery milestones related to population activity recovery. Using data from 2017 Hurricane Harvey in Harris County (Texas), the study specifies four critical post-disaster recovery milestones and calculates quantitative measurements of the length of time between the end of a hazard event and when the spatial areas (census tracts) reached these milestones based on fluctuations in visits to essential and non-essential facilities, and essential and non-essential credit card transactions. Accordingly, an integrated recovery metric is created for an overall measurement of each spatial area's recovery progression. Exploratory statistical analyses were conducted to examine whether variations in community recovery progression in achieving the critical milestones is correlated to its flood status, socioeconomic characteristics, and demographic composition. Finally, the extent of spatial inequality is examined. The results show the presence of moderate spatial inequality in population activity recovery in Hurricane Harvey, based upon which the inequality of recovery is measured. Results of this study can benefit post-disaster recovery resource allocation as well as improve community resilience towards future natural hazards.

## 1. Introduction

Post-disaster community recovery comprises short-term recovery, such as debris removal and providing essential supplies to residents (Phillips, 2009; Webb et al., 2002), and long-term recovery, such as reconstruction of the built environment and infrastructure and re-planning for future economic development (Rubin, 2009).

This subjective, task-based approach to community recovery characterization in the literature is problematic in terms of the lack of specification of critical recovery milestones, as well as an absence of quantitative data-driven measures corresponding to milestones to enable proactive monitoring to inform recovery implementation and resource allocation. Hence, departing from the standard approaches for recovery stage designations, this study proposes determination of milestones based on patterns of population activity fluctuations. Critical recovery milestones are defined as times at which important functionality of the built environment, life activities of humans, and economic activities of businesses return to their pre-disaster level (Coleman et al., 2022; Ma et al., 2022; Yuan et al., 2022). Accordingly, the recovery trajectory captures the sequence and duration of different critical recovery milestones. The specification and characterization of these critical community recovery milestones enables examination of the trajectory of recovery for different sub-populations to reveal recovery inequality (Cutter et al., 2008; Fan et al., 2022).

To this end, the goal of this study is to specify four critical recovery milestones related to population activity recovery and to provide quantitative measurements of these critical milestones using location-based population visits to points of interest (POIs) and credit card transaction data. Using data related to the 2017 Hurricane Harvey in Harris County (Houston metropolitan area), we examine neighborhood-level recovery patterns. Accordingly, we compile the duration of each critical recovery milestone to calculate an integrated population activity measure for each neighborhood. Finally, the social and spatial inequalities in recovery durations of population activities are evaluated.

## 2. Background

2.1. Post-disaster recovery monitoring

Fast and equitable recovery of all affected populations are integral components of urban resilience. By monitoring population activity, the trajectory of recovery across affected communities can be inferred. However, the current approaches to recovery evaluation and assessments have major limitations in terms of time lags incurred by placing the burden of data collection on affected populations in the form of surveys (D. Brown et al., 2008; Cutter et al., 2008; Hettige et al., 2018). In the post-disaster settings, surveys focus mainly on capturing the population's recovery progress, in the short- and long-term (Coleman et al., 2020; Martín et al., 2020). Monitoring recovery using surveys has major limitations. First, surveys put the burden of data collection (completing the surveys) on impacted populations, a method with low response rates. Second, the time-consuming process of collecting representative samples makes the longitudinal collection of data difficult. Most importantly, survey data is not timely: by the time surveys insights are available, the state of recovery has evolved significantly. These limitations of survey-based approaches to monitoring recovery are some of the reasons for inefficient prioritization and implementation of recovery resource allocation.

2.2. Emerging data for community resilience and recovery assessment

The emergence of passive geo-referenced data collected by human sensors is viewed as a potential solution to bypass the disadvantages of traditional data for examining recovery. With the rapid development of smartphones and geo-locating services, an individual's daily activities and movements leave digital footprints (Goodchild, 2007). Researchers have used these digital footprints to understand human interactions with the built environment. During disasters, the timeliness and low-cost of these datasets can reflect near real-time responses from individuals. As an efficient and effective data source, researchers have used those passively collected human movements datasets in disaster management. For example, geotagged Twitter posts are used to understand evacuation decisions and evacuation destination choices (Jiang et al., 2021; Jiang, Li, & Cutter, 2019; Martín et al., 2017). In addition, Yabe et al. (2021) examined the recovery of Puerto Rico after Hurricane Irma and Hurricane Maria based on human mobility patterns. Martín et al. (2020) used Twitter posts to examine local community recovery after Hurricane Maria in Puerto Rico, assuming that when people were able to tweet, they had basic access to electricity and telecommunication was recovered in the region. Yuan et al. (2021) measured recovery in Harris County, Texas, after Hurricane Harvey based on credit card transaction data, quantifying the recovery process of different business sectors and their disparities across the space. Podesta et al. (2021) and Coleman et al. (2022) examined visitation patterns to different types of points-of-interest and identified lifestyle changes after hurricanes. Lee et al. (2022) used human mobility data to investigate the post-disaster evacuation return movements after Hurricane Harvey. Farahmand et al. (2022) examined how human activities were disrupted by inundation during hurricanes. These recovery studies using big data utilized only a single data source, which can capture only a single perspective of the recovery progress. For example, social media-based analysis captures only a small portion of the population, so the representativeness issue remains unsolved (Jiang, Li, & Ye, 2019; Malik et al., 2015). Recovery measurements based on visitation patterns can capture the human movements after disaster but neglect local economic activities. To overcome these drawbacks, this study uses both human movement-based visitation patterns and credit card transaction datasets to derive an integrated measure of population activity recovery at the neighborhood level. Several studies have leveraged the strengths of multiple data source; for example, Yuan, Yang, et al. (2021) integrated social media, transportation, human mobility, and transaction datasets to assess damages after Hurricane Harvey.

### 3. Data and Methodology

3.1 Data Types

This study used two main data sources: location-based mobility data and credit card transaction data. Both credit card transactions and location-based visitation patterns capture the combined effects of infrastructure disruptions, household-level impacts, and business impacts. Thus, the specification of critical recovery milestones of population activities based on these datasets could provide important insights about the unfolding of post-disaster recovery.

Credit card transaction data is provided by SafeGraph, a company providing location-based datasets while protecting users' privacy by de-identifying data. This transaction dataset includes anonymized transaction data aggregated by Zip code. Each record of this dataset contains information about the merchant type, cardholder's registered Zip code, and monetary amount of the transaction. Let $m_{z,e/ne,d}$ be the total transaction amount, where $z$ represents the Zip code area, $e/ne$ indicates whether the merchant type is essential ($e$) or non-essential ($ne$), and $d$ is the date. In other words, $m_{z,e/ne,d}$ is an aggregated value of the amount paid to a certain merchant type by cardholders from a given Zip code area for a given day. Since the transaction data records are at Zip code level, we merged Zip code level to census block group (CBG)-level. In Harris County, most Zip code areas contain multiple CBGs; thus, we could break a Zip code area into the corresponding CBGs.

Location-based mobility data used in this study is provided by Spectus, a location intelligence platform that collects anonymous human mobility data from mobile apps. This dataset includes time of trip; the residential location at census block group level, which is usually the origin of a trip; and the destination of a trip as a point-of-interest. Based on this information, we processed and aggregated the data for a given day to determine the number of trips occurring from a given CBG to a POI. Let $n_{c,e/n,dt}$ be the number of trips, where $c$ represents the origin CBG, marks whether the destination POI is an essential ($e$) or non-essential ($ne$) location, and $d$ is the date. Essential or non-essential POI types are discussed in the following section.

3.2 Study area and context

Hurricane Harvey was a Category 4 hurricane that formed in August 2017 in the Gulf of Mexico and made landfall in the Houston metropolitan area on August 27, 2017 (hereafter 8/27). Hurricane Harvey's extreme precipitation marked a record flooding event in the Houston metropolitan area. It is responsible for 68 direct deaths and more than $125 billion in damage. Harvey is by far the second costliest hurricane in US history, second only to Hurricane Katrina in 2005 (Blake & Zelinsky, 2018). In this study, we focused on Harris County, within which lies the majority of the Houston metropolitan area. Harris County had an estimated population of 4.5 million in 2017, ranking it as the third most populous county in the United States. During Hurricane Harvey, Harris County suffered catastrophic flooding, as 9 out of 19 river gauges reached record high flood. 36 direct deaths recorded in Harris County and more than 160,000 structures were damaged by floods (Blake & Zelinsky, 2018). Hence, the study of Harris County during Hurricane Harvey provides a suitable testbed for evaluating the critical recovery milestones and measures proposed here.

3.3 Methods

Figure 1 shows the overview of the research workflow. The credit card transaction dataset includes credit card use behavior (amount of transaction, type of transaction, date of transaction, and Zip code of card holders). Location-based trip data directly records an individual's travel origin and destination (visitation to POIs). We first categorized whether the destination is an essential service or a non-essential service. Then we calculated four recovery measurements using both datasets and service types. Finally, we created an integrated metric to measure the overall recovery condition based on the four measurements calculated in the previous step.

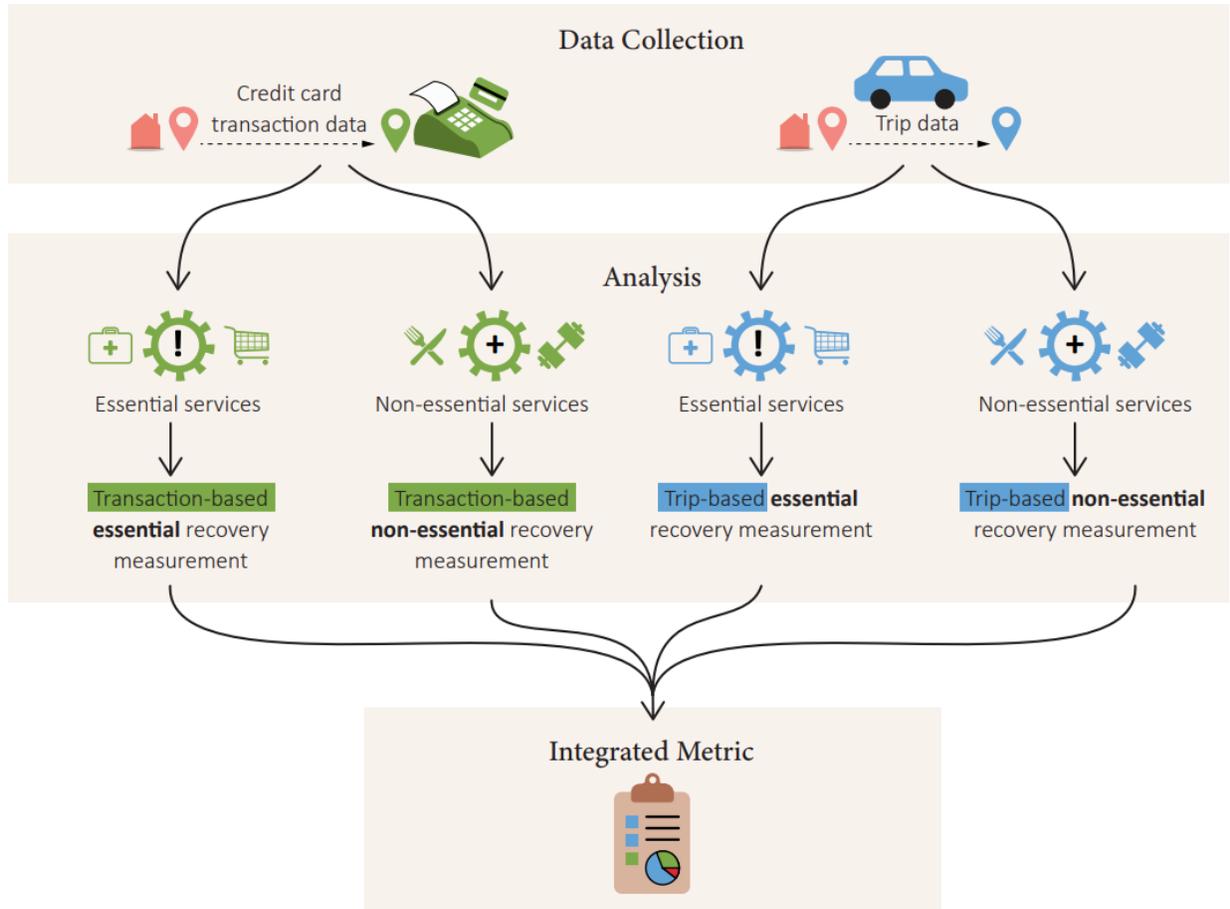

Figure 1. Overall workflow for quantitative recovery measurement

3.3.1 Essential and non-essential trips and transactions

Both the transaction dataset and the trip dataset include destination information, either a merchant, where the transaction occurred, or a POI, where the trip ended. We first categorized merchants and POIs into essential or non-essential, based on the type of services provided. Essential businesses provide services serving the general public's basic living necessities, which include, but are not limited to, gas stations, grocery stores, healthcare facilities, and utility services (Labiner et al., 2010). Examples of non-essential businesses include retail stores, restaurants, self-care stores, and recreational facilities. In the transaction dataset, the merchant type for each transaction is recorded, and thus, we categorized each transaction into essential and non-essential. For each day, we summarized the amount of money spent from an origin Zip code area at essential services and non-essential services. In the trip dataset, we categorized each trip into an essential trip or a non-essential trip, based on whether the destination POI provides essential services or non-essential services based on the North American Industry Classification System (NAICS) codes of POIs (Coleman et al., 2022).

In both essential and non-essential groups, we selected four types of services from each category for analysis. In the essential group, we selected drug store, healthcare, grocery, and utilities (electric, gas, water, and sanitary). In the non-essential group, we selected self-care, retail, recreation, and restaurants. We assigned weights to each type of service based on the number of stores/locations existing in the whole dataset. For example, 94.7% of essential services in the dataset are grocery stores. Therefore, the weight assigned to grocery store is 94.7%. Table 1 shows the weights of each selected service type.

Table 1. Service types and their weights

| Essential services | | Non-essential services | |
|---|---|---|---|
| Type | Weight (%) | Type | Weight (%) |
| Drug store | 5.0 | Self-care | 0.5 |
| Healthcare | 0.01 | Retail | 28.8 |
| Grocery | 94.7 | Recreation | 7.6 |
| Utilities | 0.2 | Restaurant | 63.1 |

After this step, we, then, summarized a weighted amount of money spent in essential and non-essential groups using Eq. 1:

$$Weighted\ measurement = \sum w_i\ m_i \qquad \text{Eq. 1}$$

where $w_i$ is the weight of service type $i$. For transaction data, $m_i$ is the total amount of money spent at service type $i$ for credit card transaction data, and for trip data, $m_i$ is the number of trips ended at service type $i$. Accordingly, we calculated the weighted values of transactions and trips for two categories of essential and non-essential services in an aggregated manner.

3.3.2 Baseline and recovery calculation

Figure 2 illustrates the schematic representation of the disruption and recovery periods following a disaster. Pre-disaster activities are considered as the steady state, and thus are used in this study as baseline values. When Hurricane Harvey hit Harris County, it disrupted both residents' travel and business transactions. During the recovery period, flood water receded and impacts dissipated, the trips and transactions gradually returned to near steady-state levels.

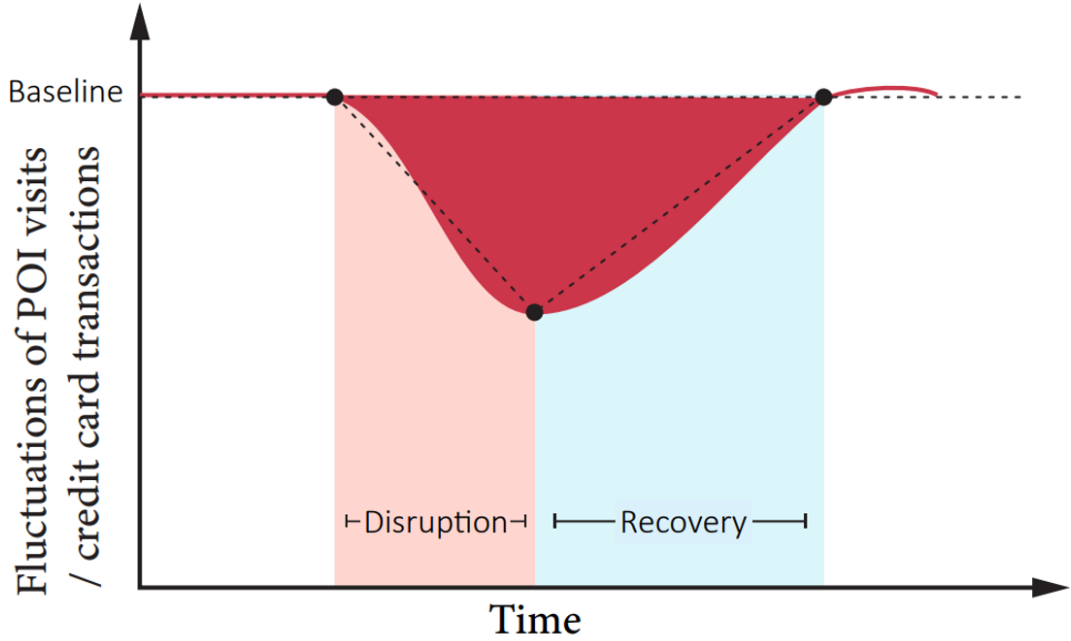

Figure 2, Disruption and recovery curve following a disaster

Since Hurricane Harvey hit Harris County on 8/27, we chose the baseline period as daily average transactions between August 1 and August 21, 2017, eliminating the days immediately before hurricane's arrival may include irregular traveling and shopping patterns due to hurricane preparation. We calculated the average daily trips to essential and non-essential POIs starting from each CBG during this 21-day window and used this average trip length as the baseline for trips in this CBG for both essential and non-essential services using the following Eq. 2:

$$\overline{n_{c,e/ne}} = \frac{1}{21}\sum_{d=1}^{21} n_{c,e/ne,d} \qquad \text{Eq. 2}$$

Similarly, we calculated the average daily amount of money spent during this 21-day window and used this average amount as our credit card transaction baseline for essential and non-essential services using Eq. 3:

$$\overline{m_{z,e/ne}} = \frac{1}{21}\sum_{d=1}^{21} m_{z,e/ne,d} \qquad \text{Eq. 3}$$

After this step, we have four baselines for each CBG: credit card transaction for essential services ($m_{z,e}$), credit card transaction for non-essential services ($m_{z,ne}$), trips to essential services ($n_{c,e}$), and trips to non-essential services ($n_{c,ne}$).

For each day after 8/27, we first divided transactions and trips into essential and non-essential groups. Then we applied a smoothed average by calculating a 7-day moving average that included 3 days before the focal date and 3 days after the focal date to ease weekly fluctuations.

Based on data types (transactions or trips) and whether they are associated with essential or non-essential services, we then computed the percentage change of transaction amounts and trip counts using the following two equations, respectively:

$$Transaction\ change = \frac{m_{z,e/ne,d} - \overline{m_{z,e/ne}}}{\overline{m_{z,e/ne}}} \qquad \text{Eq. 4}$$

$$Trip\ change = \frac{n_{c,e/ne,d} - \overline{n_{c,e/ne}}}{\overline{n_{c,e/ne}}}$$

Daily essential or non-essential transaction amounts and trip counts are compared with their corresponding baseline values to calculate the change percentage in each day following the landfall of Harvey.

3.3.3 Recovery time calculation

In this study, we defined a population activity in a spatial area as "recovered" when we observed three days of post-disaster activities reached 90% of the baseline values. Accordingly, recovery duration is calculated using Eq. 5:

$$t = d_0 - d_n \qquad \text{Eq. 5}$$

where, $d_0$ is the day when hazard occurred and $d_n$ is the third day in which the post-disaster activities were observed to reach 90% of the baseline.

Since we have four different categories of recovery milestones, each spatial area (CBG) has four measurements for recovery milestones. For example, when we observed that a CBG's transaction amounts for essential services reached 90% of the baseline for 3 days, we considered the third day as $d_n$ for this region's essential services recovery, and the number of days between $d_0$ and $d_n$ is the transaction-based essential recovery milestone duration for this CBG. Since we are using trip-based and transaction-based measurements for both essential and non-essential services, we have four different recovery measurements for each census tract, all measured by number of days.

3.3.4 Integrated population activity recovery metric

After the previous calculation, we obtained four measurements of recovery milestones for each CBG: trip-based essential, trip-based non-essential, transaction-based essential, transaction-based non-essential. In this step, we created an integrated metric to combine these four measurements that provides an overall measurement of quickly fast population activities in a neighborhood recovered. This includes two main steps: the first step is normalization; the second step is average.

First, we used the min-max scaling method to normalize each measurement using the following equation:

$$t_{norm} = \frac{t_i - t_{min}}{t_{max} - t_{min}} \qquad \text{Eq. 6}$$

where $t_{norm}$ is the normalized value for the recovery duration; $t_i$ is the observed recovery duration for region $i$; $t_{min}$ and $t_{max}$ are the minimum and maximum duration of recovery, respectively. This step projected all the four measurements onto a scale between 0 and 1.

Secondly, we take average of these four normalized measurements and define this averaged measurement as our integrated metric:

$$\text{Integrated metric} = \frac{\sum \text{Normalized Metric}}{4} \qquad \text{Eq. 7.}$$

After calculating the integrated metric, we divided all the values by quantile. The lowest 25% are labeled as early recovery; 25% to 50% are mild recovery; 50% to 75% are late recovery; and those above 75% are delayed recovery.

### 3.3.5 Influencing factors

Existing studies have pointed out that recovery is not consistent across all the neighborhoods in impacted areas and that different demographic and socioeconomic characteristics affect a community's recovery progress (Cutter et al., 2006; McDonnell et al., 1995). To understand social disparities and inequality in recovery from Hurricane Harvey, we examined the variations in the integrated recovery metric with respect to flood status and socioeconomic status of CBGs.

We chose three variables that may be associated with the community's recovery progress to further examine these relationships: flood status, minority percentage, and per capita income. Minority percentage and per capita income datasets were obtained from the US Census Bureau. For these two socioeconomic variables, we used the median value as the categorizing criteria. Those CBGs with an above-median value were marked as 1; those with a lower-than-median value were marked as 0. For flood status, we obtained flooded area data for Hurricane Harvey from the Federal Emergency Management Agency (FEME). Then we calculated the flooded area percentage for each CBG. We used the median flooded percentage as the criterion. Those CBGs that have more-than-median flooded areas are marked 1, and those with fewer-than-median flooded areas are marked 0.

To test the correlation between recovery speed and these three variables, we conducted correlation tests to explore their relationships. In addition, chi-square tests were conducted with the null hypothesis stating that population activity recovery speed in Harris County at CBG-level has no relationship with flood status and sociodemographic attributes of CBGs.

## 4. Results and Discussion

### 4.1 Recovery measurements based on trips and transactions

Figure 3 shows the recovery duration for the four recovery milestones. We mapped Zip code areas to CBGs to unify the spatial units in this study. In Harris County, the majority of Zip code areas contain several CBGs. If a CBG is fully inside a Zip code area, this CBG takes the value for the overarching Zip code area. If a CBG is divided into more than one Zip code, we assigned the CBG's value as the Zip code with the largest area.

### 4.1.1 Trip-based essential activity recovery

Figures 3(a) and 3(b) illustrate the recovery duration for essential and non-essential trips for each CBG in Harris County. We found that 606 CBGs having trip-based essential activity recovered within 1.5 weeks, which is about 28.3% of all the CBGs. In 418 CBGs, trip-based essential activity recovery took between 1.5 to 2.5 weeks. A total of 218 CBGs recovered between 2.5 to 3.5 weeks, based on trip records to essential services. And 132 CBGs observed with trip-based essential activity recovered between 3.5 to 4.5 weeks. The trip-based essential activity recovery, however, took more than 4.5 weeks for 770 CBGs, more than 35% of all the CBGs. Figure 4(a) illustrates the distribution of the numbers of CBGs corresponding to the recovery time. For trip-based non-essential activity recovery, only 492 CBGs are found recovered within 1.5 weeks, and 375 CBGs recovered between 1.5 to 2.5 weeks. In 217 CBGs, trip-based non-essential activity recovery took between 2.5 to 3.5 weeks. The remaining 152 CBGs showed that trip-based non-essential activity recovered between 3.5 to 4.5 weeks. However, 908 CBGs recovered more than 4.5 weeks, which is more than 42% of all the CBGs (Figure 4(b)).

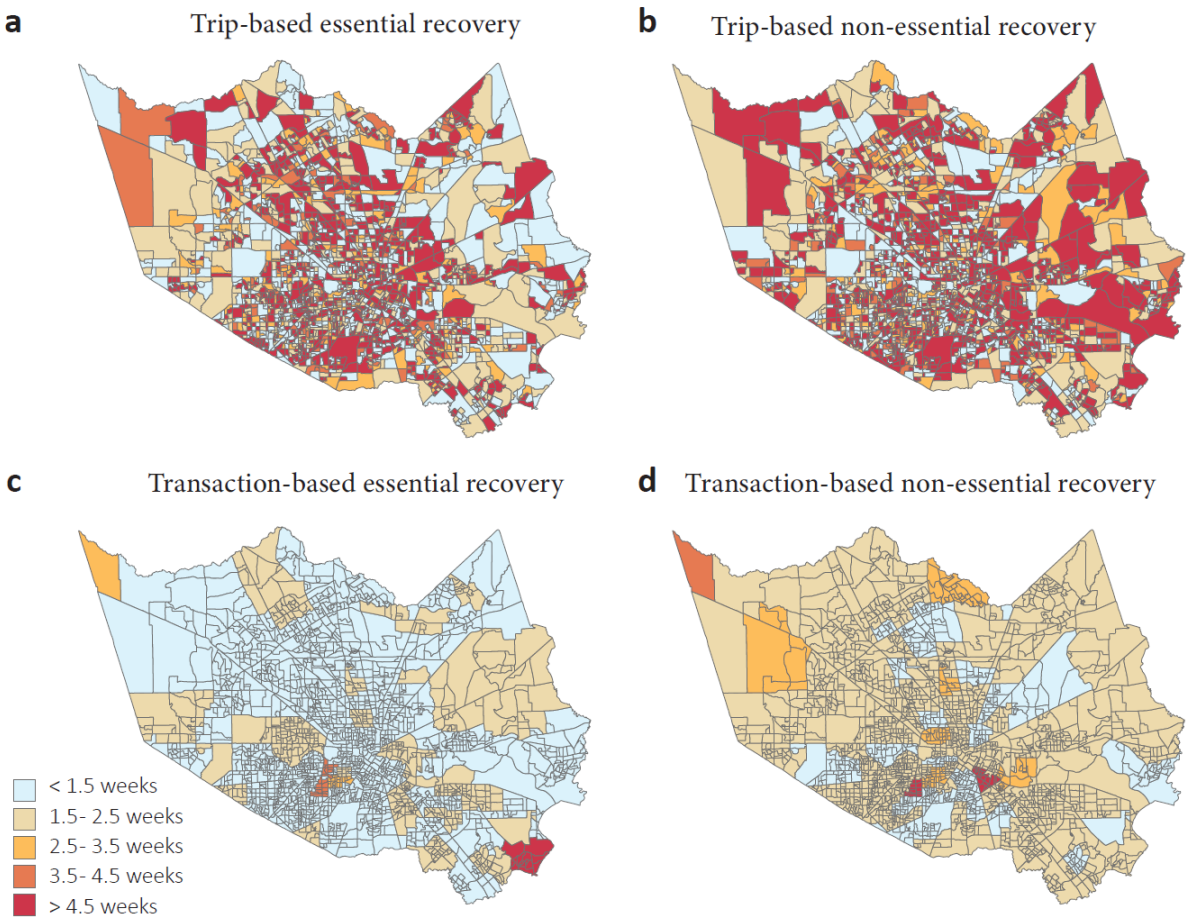

Figure 3. Recovery duration for (a) trip-based essential recovery; (b) trip-based non-essential recovery; (c) transaction-based essential recovery; and (d) transaction-based non-essential recovery.

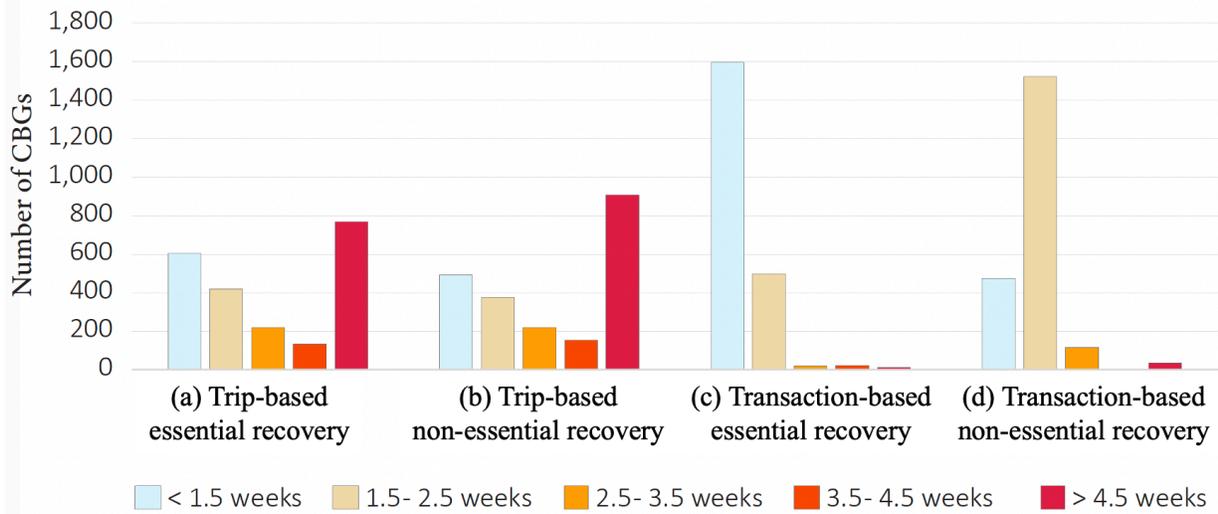

Figure 4. Distribution of the number of CBGs recovered with corresponding recovery time.

Among these 2,144 CBGs with mobility data, 859 CBGs were observed to have trip-based essential activity recovery duration shorter than trip-based non-essential activity recovery duration. In 865 CBGs, essential and non-essential activity recovery duration required roughly the same time. In the other 420 CBGs, non-essential activity recovery duration was found to be shorter than essential activity recovery duration based on trip records (Figure 5(a)). Figure 6 demonstrates the comparison between essential activity recovery duration and non-essential activity recovery duration. In general, we found that CBGs on the east side of Harris County recovered essential activities earlier than non-essential activities. We also conducted Moran's test to examine spatial patterns. Moran's I had z-value of 11.916 for trip-based essential activity recovery and 8.133 for trip-based non-essential activity recovery. The p-values for both are smaller than 0.001, indicating the existence of significant spatial clusters. This result suggests the presence of spatial effects in recovery patterns of communities.

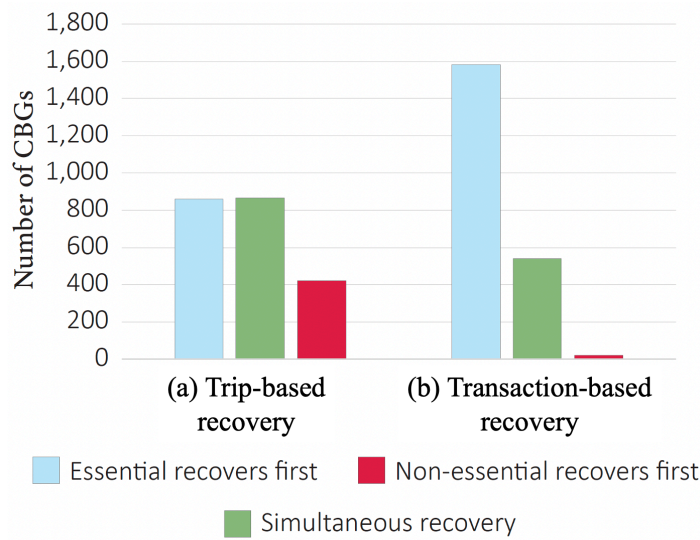

Figure 5. Comparison of recovery time between essential activity and non-essential activity.

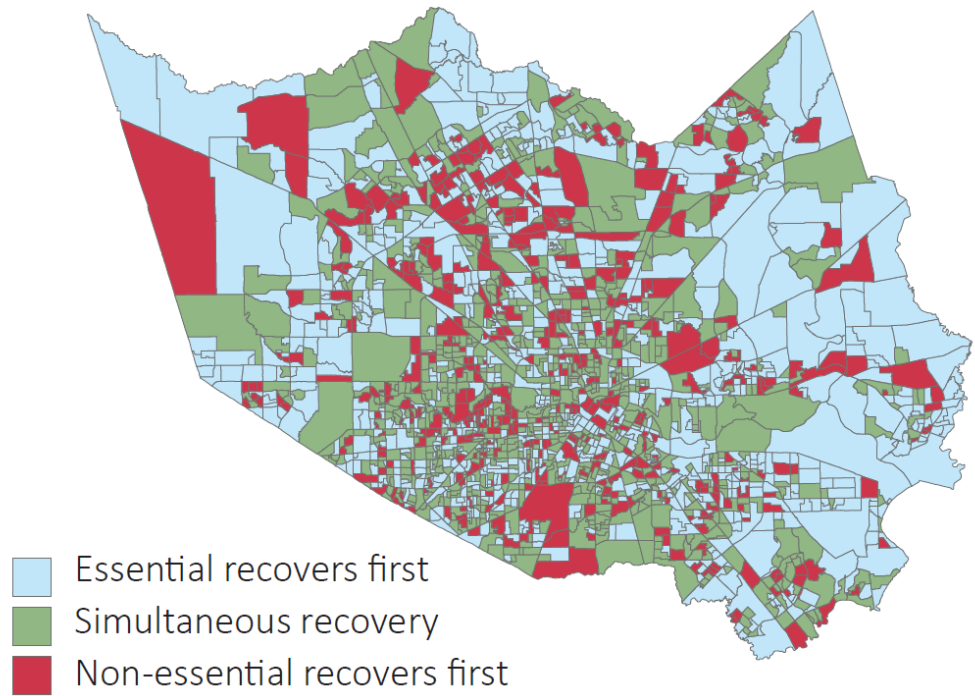

Figure 6. Comparison between trip-based essential and non-essential activity recovery.

4.1.2 Transaction-based essential and non-essential activity recovery

Figure 3(c) and figure 3(d) depict transaction-based essential and non-essential activity recovery. In general, we found that transaction-based essential activities recovered fastest after Hurricane Harvey, as 1,596 out of 2,144 CBGs recovered in less than 1.5 weeks, which included 74.4% of all CBGs. On the other hand, we also found 11 CBGs took more than 4.5 weeks to recover, all of which are clustered at the southeast corner of the county near Trinity Bay, which includes complicated river networks and wetlands. During Hurricane Harvey, heavy precipitation caused Harris County watersheds to overflow; neighborhoods in these areas suffered more extensive damage than other areas of the county. Analysis revealed that 496 CBGs show transaction-based essential activity recovery took some time between 1.5 to 2.5 weeks; 19 CBGs recovered between 2.5 to 3.5 weeks; and other 22 CBGs show transaction-based essential recovery took 3.5 to 4.5 weeks (Figure 4(c)). In general, non-essential activities recovered slightly later than essential activities based on the transaction records. Average non-essential activity recovery time was 1.8 weeks, compared to essential activity recovered in 1.4 weeks. Based on the transaction-based non-essential activity data, 473 CBGs recovered in less than 1.5 weeks, which includes 22% of all CBGs. The majority of areas (1,522 CBGs) recovered between 1.5 weeks to 2.5 weeks. Also, we found recovery took more than 4.5 weeks in 33 CBGs (Figure 4(d)). One cluster of these neighborhoods is located at the east boundary of University of Houston, including Greater Eastwood, Gulfgate/Pine Valley, and parts of Lawndale/Wayside. The other cluster of delayed non-essential recovery neighborhoods is located in Bellaire, where about one-third houses were damaged (Jones, 2017).

Figure 7 shows the comparison of recovery time between essential and non-essential businesses based on transaction records. We can see that in 1,582 CBGs, out of 2,144 CBGs in total, essential activities recovered earlier than non-essential activities based on transaction records. In 540 CBGs, transaction-based essential and non-essential activity recovery took about the same time. In other 22 CBGs, non-essential businesses recovered earlier than essential businesses by more than a week. (Figure 5(b)) To test whether such patterns are spatially clustered, we conducted Moran's I test and examined the spatial autocorrelation. Results for Moran's I show that the z-score for essential transaction activity recovery is 69.98 for essential transaction activity recovery and 70.08 for non-essential recovery. The p-values for both tests are smaller than 0.001, indicating that for both transaction-based essential and non-essential activity recovery milestones, significant spatial autocorrelations are present.

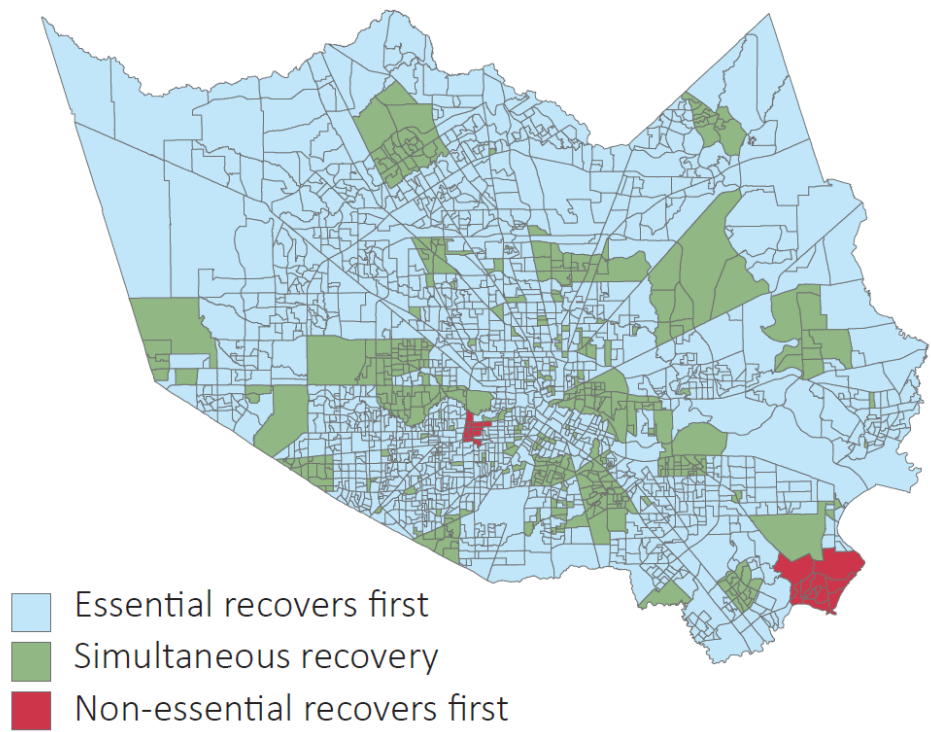

Figure 7. Comparison between transaction-based essential and non-essential recovery.

Comparing recovery based on transaction and trips data, transaction-based activities demonstrate faster recovery than trip-based activity recovery. The majority of CBGs are found to be recovered within 1.5 weeks based on transaction records. One potential explanation is that residents spent more per transaction on needed supplies after the hurricane. If roads were flooded, residents had to reduce the total number of trips, but they still need to visit places such as grocery stores. They may be likely to purchase all necessary supplies in one trip to avoid the inconvenience of multiple trips. Therefore, when we consider the total amount spent within a CBG, this value shows a faster recovery speed than the number of trips a CBG has made.

4.2 Integrated metric for population activity recovery

Figure 8 shows the integrated population activity recovery metric calculated based on the method presented in section 3.3.4. Based upon scaled measurements, the integrated metric value for a CBG, between 0 and 1, indicates the speed of recovery. The smaller this integrated metric value, the shorter the time to recovery. Among all the 2,144 CBGs with sufficient data in Harris County, the integrated metric ranges from 0.004 to 0.883, with a mean of 0.220 and a standard deviation of 0.164. We further categorized CBGs based on their integrated metric percentile. Early recovery groups contain CBGs with integrated metrics less than 0.08, which includes 25% of the fastest-recovered CBGs. Mild recovery group includes recovery speed between 25% to 50%, with the integrated metric value from 0.08 and 0.18. CBGs with an integrated metric value between 0.18 and 0.32 are grouped as late recovery, containing those ranked from 50% to 75%. The rest of CBGs are grouped as delayed recovery for those with an integrated metric value larger than 0.32. As shown in Figure 5, CBGs having late or delayed recovery are clustered in the downtown Houston area. CBGs in the early recovery group are in the west and north area, mostly outside the Texas State Highway 8 beltway. Neighborhoods with delayed recovery are located along the major rivers traversing the Houston metropolitan area. Most of these neighborhoods are located along Buffalo Bayou, White Oak Bayou, and Sims Bayou. When rivers were flooded due to the extensive precipitation caused by Hurricane Harvey, neighborhoods along these rivers suffered extensive damages from the overflow (Abouzir, 2020). Comparing the four measurements of recovery, we found that trip activities have a greater influence on the integrated metric. These neighborhoods with delayed recovery are also found to have more than 4.5 weeks of trip-based activity recovery. Moran's I result for this integrated metric value shows that z-score is 17.758 and the p-value is smaller than 0.001. This means the integrated population activity recovery metric values exhibit significant spatial autocorrelation; in other words, neighbor communities' recovery progress affected each other. The role of spatial effects in diffusion of recovery among neighborhoods can be further studied in future research (Liu & Mostafavi, 2022).

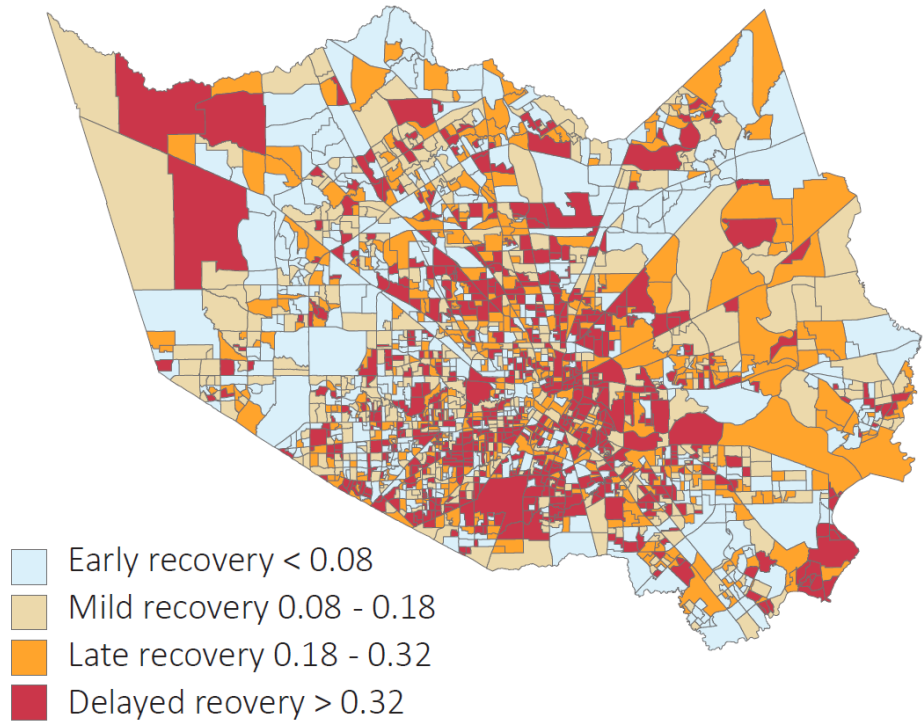

Figure 8. Integrated metric for recovery measurement.

4.3 Influencing factors of population activity recovery

In this section, we investigate whether population activity recovery speed was correlated to socioeconomic and demographic characteristics, as well as to flood status of each neighborhood. Chi-squared tests were conducted to test the null hypothesis that no relationships exist on the categorical variables. Specifically, the categorical variables we used in this study were minority percentage, per capita income, and flood status. As mentioned in Section 3.3.5, for each CBG, we categorized minority percentage, per capita income, and flood status into high-level, marked as 1, and low-level, marked as 0. In addition, we also categorized recovery status using integrated metric value. We grouped CBGs that recovered more quickly than the median value (early and mile recovery categories) as fast recovery. Those in the late recovery and delayed recovery categories were grouped as slow recovery. Chi-squared tests were conducted to test whether fast or slow recovery has relationships with minority percentage, per capita income, and flood status.

Table 2 shows the p-values for the three chi-squared tests. The p-values for per capita income and minority percentage were smaller than 0.001, meaning we can reject the null hypothesis. However, the p-value for recovery speed and flood status is 0.870, meaning we don't have enough evidence to reject the null hypothesis. Based on these chi-squared tests, we can conclude that the duration of a CBG's recovery does not have a significant relationship with its flood status, but the recovery progress shows a relationship with its minority percentage and per capita income.

Table 2. p-values for chi-squared tests.

| Variables | p-value |
|---|---|
| Per capita income | <0.001 |
| Minority percentage | <0.001 |
| Flood status | 0.870 |

For flood status, we simply categorized flood severity based on flooded areas within each CBG. Due to the data limitations, we could not explore the flooded area to further investigate this relationship. It is possible that flooded areas were plains or vacant lands, rather than roads or infrastructures. Therefore, some places may report a large, flooded area but most trips were unaffected. To further examine this, a higher-resolution dataset related to road flooding situations would be needed.

Despite the identification of social inequality in the speed of population activity recovery, we found that some low-income neighborhoods recovered early. This may be related to their pre-disaster lifestyles. Low-income households are more likely to spend a large portion of their income on essential supplies, such as food, water, gas, and utilities. These essential supplies were grouped as essential activities in our study. low-income families show very similar patterns when comparing pre-disaster and post-disaster travel behaviors and consumption behaviors. In fact, after disasters, the stores frequented residents of low-income CBGs in pre-disaster may run out of essential supplies, necessitating travel to a more distant store to purchase essential supplies. From a data perspective, those neighborhoods may show an increase in the number of trips, a longer travel distance, and an increase in money spent on essential supplies (Esmalian et al., 2022).

On the other hand, households with higher income could afford long-distance evacuation to a different city. These evacuated people may return after they have confirmed local infrastructures were fully rebuilt and stores were stocked. From a data perspective, those who evacuated to locations outside of Harris County were beyond our data collection range, and thus, their home community presented a slower population activity recovery pattern because those evacuated people may return after surrounding areas were recovered and infrastructure services were restored.

4.4 Spatial inequality in population activity recovery

We calculated the Gini index, used in economics as a measure of statistical dispersion, to measure the spatial inequality of the distribution of the integrated population activity recovery metric. The Gini index ranges between 0, meaning perfect equality, and 1, meaning perfect inequality. The Gini index can measure spatial inequality with respect to different attributes (Ceriani & Verme, 2012; US Census Bureau, 2021). Outside of economics, the Gini index is also used in health resources allocation (M. C. Brown, 1994; Erdenee et al., 2017), land-use structure (Zheng et al., 2013), and algorithm decision making (Sundhari, 2011; Wang & Xu, 2008). Considering that each CBG has different built environment features and has suffered different levels of damages from the Hurricane Harvey, it is normal for population activities of CBGs to recover at different speeds. In an ideal situation with perfect equality, the integrated recovery metric distribution would follow the red line in Figure 9. However, the actual integrated recovery metric distribution is shown as the black curve, known as Lorenz Curve for the Gini index calculation, formed with individual CBG's integrated recovery metric values. The Gini index is calculated as the difference between the ideal situation (line) and the actual distribution (Lorenz Curve). After calculation, the Gini index for population activity recovery inequality follow Hurricane Harvey is 0.41, indicating a moderate spatial inequality in population activity recovery. Such inequality can be caused by multiple factors, including demographic composition, socioeconomic status, and flood damage. Since the Gini index only provides an overall measurement for spatial inequality, further analyses are needed to understand the determinants of inequality in population activity recovery.

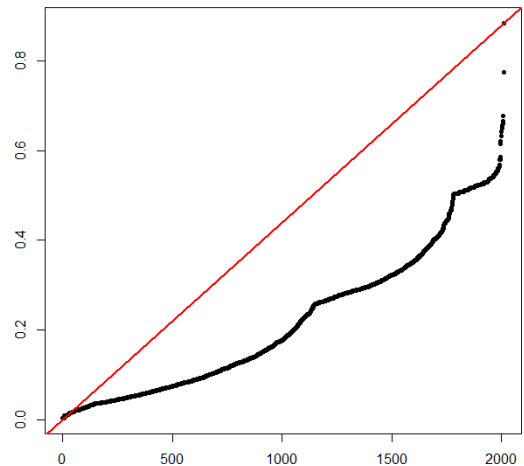

Figure 9. Lorenz curve of the integrated recovery metric distribution (black curve) comparing to the distribution with the perfect equality distribution (red line).

## 5. Concluding Remarks

This study specified and quantified four critical recovery milestones associated with population activities to facilitate better understanding and monitoring of the recovery trajectory of post-disaster recovery. These findings offer multiple significant scientific and practical contributions. The characterization and quantification of critical recovery milestones advances the understanding of ways recovery unfolds in communities following a hazard event. The measures associated with the four recovery milestones enable quantifying and monitoring the progression of recovery at fine spatial resolutions (census block group level). These measures are computed using passively sensed population activity data that are privacy compliant. This data collection method relieves the burden of data collection from impacted residents with the further advantage of near-real-time data provision. From a practical perspective, emergency managers, public officials, and decision-makers can utilize the findings of this study to proactively monitor the progression of recovery in different neighborhoods and identify areas which show delayed recovery to better allocate recovery resources. Data-driven recovery monitoring would significantly improve the efficiency of resource allocation and prioritization which in turn accelerates the recovery of the entire community and reduce the overall impacts of disasters on people. These advancements open avenues for future research to empirically further characterize the spatial and temporal progress of post-disaster recovery for better evaluating critical recovery milestones and their spatiotemporal interdependence, as well as the determinants of variations in critical recovery milestones across a community.

This study also has limitations inherent in the nature of the datasets used. Since we used human mobility and credit card transaction datasets as the main data sources in this study, a limitation of this study lies in the representativeness of the sampled data. Although a recent report shows that in 2020 about 79 percent of Americans owned at least one credit card (Gonzalez-Garcia & Holmes, 2021), credit card transaction records still could not capture population activity behavior for a segment of a low-income population. Also, in a report by Pew Research, about 77 percent of Americans owned a smartphone (Pew Research Center, 2021). Recent studies have shown the representativeness of location-based data. However, to tackle this limitation, future studies could identify solutions to examine the extent of biases in location-based and transaction datasets and develop solutions for data de-biasing during data pre-processing and processing if data biases are identified.

### Data availability

All data were collected through a CCPA- and GDPR-compliant framework and utilized for research purposes. The data that support the findings of this study are available from Spectus Inc., but restrictions apply to the availability of these data, which were used under license for the current study. The data can be accessed upon request submitted on Spectus.ai. Other data we use in this study are all publicly available.

### Code availability

The code that supports the findings of this study is available from the corresponding author upon request.

### Acknowledgments

**This material is based in part upon work supported by Texas A&M University X-Grant 699.** The authors also would like to acknowledge the data support from Spectus, Inc. Any opinions, findings, conclusions, or recommendations expressed in this material are those of the authors and do not necessarily reflect the views of Spectus, Inc.